\documentclass[10pt,aps,prb,amsmath,amssymb,floatfix,raggedbottom,floats,superscriptaddress,twocolumn]{revtex4-1}

\usepackage{graphicx}% Include figure files
\usepackage{dcolumn}% Align table columns on decimal point
\usepackage[mathlines]{lineno}% Enable numbering of text and display math
\usepackage{amsthm,bm}
\usepackage{mathrsfs}
\DeclareMathAlphabet{\mathscrbf}{OMS}{mdugm}{b}{n}
\usepackage{etoolbox}
\usepackage{mathtools}
\usepackage{xparse}
\NewDocumentCommand \BiG {m O{} O{}}{
        \IfNoValueTF{ #2}
        {
                \IfNoValueTF{ #3}
                {
                        \big{.#1}
                }
                {
                        \big{.#1}^{ #3 }
                }
        }
        {
                \IfNoValueTF{ #3}
                {

                        \big{.#1}_{ #2}
                }
                {
                        \big{.#1}_{ #2}^{ #3}
                }
        }
}

\DeclareMathOperator{\vctrz}{vec}
\DeclareMathOperator{\Diag}{Diag}

\DeclareMathOperator{\Tr}{Tr}
\usepackage[dvipsnames]{xcolor}
\usepackage[normalem]{ulem}

\begin{document}

\title[]{Efficient determination of the Markovian time-evolution towards a steady-state\\ of a
         complex open quantum system}

\author{Thorsteinn H.\ Jonsson}
\affiliation{Science Institute, University of Iceland, Dunhaga 3, IS-107 Reykjavik, Iceland}%
\author{Andrei Manolescu}
\email{manoles@ru.is}
\affiliation{School of Science and Engineering, Reykjavik University, Menntavegur 
             1, IS-101 Reykjavik, Iceland}%
\author{Hsi-Sheng Goan}
\affiliation{Department of Physics and Center for Theoretical Sciences, National Taiwan University, 
             Taipei 10617, Taiwan}
\affiliation{Center for Quantum Science and Engineering, 
             National Taiwan University, Taipei 10617, Taiwan}
\email{goan@phys.ntu.edu.tw}
\author{Nzar Rauf Abdullah}
\affiliation{Physics Department, College of Science, 
             University of Sulaimani, Kurdistan Region, Iraq}
\author{Anna Sitek}
\affiliation{Department of Theoretical Physics, Faculty of Fundamental Problems of Technology, 
            Wroc{\l}aw University of Technology, 50-370 Wroc{\l}aw, Poland}
\email{anna.sitek@pwr.edu.pl}
\author{Chi-Shung Tang}
\affiliation{Department of Mechanical Engineering, National United University, Miaoli 36003, Taiwan}
\email{cstang@nuu.edu.tw}
\author{Vidar Gudmundsson}%
 \email{vidar@hi.is}
\affiliation{Science Institute, University of Iceland, Dunhaga 3, IS-107 Reykjavik, Iceland}%

\date{\today}%,
\begin{abstract}
Master equations are commonly used to describe time evolution of open systems. 
We introduce a general computationally efficient method for calculating a Markovian solution 
of the Nakajima-Zwanzig generalized master equation. 
We do so for a time-dependent transport of interacting electrons through 
a complex nano scale system in a photon cavity. The central system, described by 120 many-body
states in a Fock space, is weakly coupled to the external leads. 
The efficiency of the approach allows us to place the bias window defined by the external
leads high into the many-body spectrum of the cavity photon-dressed states of the central 
system revealing a cascade of intermediate transitions as the system relaxes to a steady
state. The very diverse relaxation times present in the open system, reflecting radiative 
or non-radiative transitions, require information about the time evolution through many orders 
of magnitude. In our approach, the generalized master equation is mapped from a many-body 
Fock space of states to a Liouville space of transitions. We show that this results in a linear 
equation which is solved exactly through an eigenvalue analysis, which supplies information on the 
steady state and the time evolution of the system.
\end{abstract}
\maketitle

\section{Introduction}
Growing interest is manifesting itself in nonequilibrium processes during
electron transport in, or through, small systems in photon 
cavities.\cite{PhysRevX.6.021014,PhysRevLett.109.077401,PhysRevLett.99.206804,Gustavsson06:076605}
Diverse aspects of nonequilibrium dynamics of quantum systems have often been
investigated using master equations of various types, tailored to the
task at hand.\cite{Haake1973,RevModPhys.47.67,PhysRevB.87.195427,PhysRevB.81.155303} 
For the description of the transient time regime, 
immediately following a ``switch-on'' of a perturbation of the environment
such as coupling to leads, or a photon/phonon bath, generalized master equations (GME) have prevailed. 
These equations can be traced back to Nakajima and Zwanzig, who derived the GME  
projecting the dynamics of the whole system onto the small central system, under the influence 
of the environment.\cite{Zwanzig60:1338,Nakajima58:948,Haake1973} 

The search for a steady state of a nonequilibrium
system can be accomplished by requiring that
the time-derivative of the density operator vanishes and
instead of the first order differential equation one is left
with a homogeneous algebraic Sylvester equation or an eigenvalue
problem in a larger space than the calculation was started in.
The eigenvalue problem is often avoided except for systems with
few dynamically active states. 

An open source toolbox has been written in Python to 
facilitate the programming of the dynamic evolution of
open systems.\cite{Johansson20131234,Johansson20121760}        
The authors focus on iterative approaches for finding the eigenvalue spectrum in Liouville space for different
reordering strategies of the Liouvillian matrix. For our system we do
not find a need for any preconditioners, instead we provide a thorough analysis of the sparse matrix
representation which results from our mathematical treatment of the model.

In a prior publication, we show how a Markovian form of the GME can be used to describe an electron transport through a
nano-scale system in a photon cavity.\cite{2016arXiv160508248G}
The GME is derived in such a way that its non-Markovian form gives an adequate description of our system
in the transient time regime, after the moment the electrons begin to enter the system in the cavity.
With current computational resources the non-Markovian approach is limited to a short time scale,
typically of hundreds of picoseconds. In the Markovian limit, the description of this system can be 
extended much further, sometimes up to 12 orders away on the time scale. 
The Markovian limit is used to attain a mapping of the description to the Liouville 
space.
In the past, authors have considered the Liouville space \(\mathfrak{L}\) for nonequilibrium dynamics of
simple systems.\cite{Weidlich71:325} By vectorizing the GME one can obtain a sparse matrix representation of
the Liouvillian in the chosen basis.\cite{PhysRevE.91.013307} 
The Liouvillian has been considered in terms of its spectral properties \cite{Nakano2010,Petrosky01032010}
and researchers have obtained a fractal description for it, sensitive to rational and irrational values similar 
to the Hofstadter's butterfly.\cite{Hofstadter76:2239}

Here, we shall elucidate in details how this can be accomplished for a 
Fock-space including 120 many-body states of a complex interacting electron-photon system in a manner appropriate for 
efficient parallel computation. Upon a closer examination of the factors totalling the Liouvillian, the Markovian
form of the GME can be explained clearly in terms of the block matrix 
factors of our sparse matrix representations.

To understand the needs for the long time
tracking of the system we will give a short introduction to the system and
subsequently introduce our methodology based on a mathematical analysis of
the structure of the GME. Subsequently, we display results showing the 
effects of the very diverse relaxation times at work in the system
as the bias window defined by the external leads is placed high in the many-body energy
spectrum of the photon-dressed electron states of the central system. We observe a cascade
of intermediate transitions, radiative or nonradiative, as the system approaches the steady state. 

The methodology is not particular to our type of a GME and should also be of
use for other problems involving master equations for complex systems.
Furthermore, the presented methodology could open up the possibility to effective
calculations of the time evolution of open systems with higher order coupling 
between the subsystems, beyond the weak coupling limit.   

\section{Theory}
\subsection{The model}
The system is described by the Hamiltonian
\begin{equation}
      H = H_S + H_{LR} + H_T,
\end{equation}
where \(H_S\) is the Hamiltonian of the central system which is weakly coupled to the 
surrounding environment, represented by the left and right leads \(H_{LR}\). The coupling of the 
central system to the leads through the contact region between these 
two systems is described by the transfer or coupling Hamiltonian 
\(H_T\).\cite{Moldoveanu07:0706.0968,Moldoveanu09:073019,Gudmundsson09:113007}

The central system consists of Coulomb interacting electrons in a finite
parabolic 2D GaAs quantum wire with length $L=150$ nm. In the transport direction it
is terminated with hard walls in the contact area at $x=\pm L/2$. The parabolic
confinement in the $y$-direction has characteristic frequency $\Omega_0=2.0$ meV.
A weak homogeneous external magnetic field along the $z$-direction is used to break
the spin-degeneracy of the electrons.    
The parabolic confinement and the external magnetic field define a length scale
$a_w=(\hbar/(m^*\Omega_w))^{1/2}$, where $\Omega_w=(\omega_c^2+\Omega_0^2)^{1/2}$,
and $\omega_c=(eB/(m^*c))$.
The central system is embedded in a rectangular photon cavity with one mode of
energy $\hbar\omega$, and a polarization in the $x$-direction for a TE$_{011}$
mode, or in the $y$-direction for a TE$_{101}$ mode. Both the paramagnetic 
and the diamagnetic part of the electron-photon interactions are accounted for.
The interactions of the system, which are the electron-electron\cite{Moldoveanu10:155442}and the 
electron-photon interactions, are treated by stepwise diagonalizations and truncation
of the applied many-body Fock-spaces.\cite{Gudmundsson:2013.305}

The coupling to the leads is described by
\begin{equation}
    H_T = \chi l(t) \int d  \mathbf{q} \left(\mathcal{T}^l(\mathbf{q})c_{\mathbf{q}}^l +
          c_{\mathbf{q}}^\dagger(\mathcal{T}^l(\mathbf{q}))^\dagger\right),
\end{equation} 
where the index $l\in{L,R}$ labels the left and right external leads.
\(\chi^l\) is a switching function, which is $0$ for $t<0$, meaning that the leads
were disconnected from the central system in the past and a connection is established
at $t=0$. \(\mathbf{q}\) is a dummy index/quantum number representing the momentum
of a state in the semi-infinite leads and its subband index. The Coulomb interaction is 
neglected between the states in the leads, but the leads are subjected to the same external
homogeneous magnetic field as the central system. 

\subsection{Time evolution}
The time evolution of the open system after the coupling to the leads can be 
described by the Liouville-von Neumann equation, given by
\begin{equation}
    \partial_t \rho = \mathcal{L}\rho .
\end{equation}
Here, \( \mathcal{L}\) is the Liouville operator, defined by the commutator
\begin{equation}
    \mathcal{L}\rho = -i/\hbar\left[H, \rho\right],
\end{equation}
and \(\rho\) is the total density operator of the system, describing the dynamics of both the leads and the central
system. We denote by $\mathbb{L}$ the corresponding solution space of density operators and call it the Liouville space.
Before the leads are coupled to the central system we assume that the density matrices are uncorrelated, i.e.\ given by
the tensor product
\begin{equation}
    \rho(t_0) = \rho_{S}(t_0)\rho_l(t_0),
\end{equation}
where \(\rho_{S}(t_0)\) and \(\rho_l(t_0)\) are the initial values of the reduced density operator for the central
system and the density operator of the leads respectively. Assuming that the leads are in equilibrium, the density
operator of the leads can be described by the grand canonical distribution.\cite{Nakajima58:948} 
For large systems, the Liouville-von Neumann equation alone is not viable for an effective calculation since the continuous state space 
of the leads is too large to handle effectively.
A way out of this dilemma was suggested by Nakajima and Zwanzig who invented a scheme 
to project the dynamics of the whole system onto the open central system.   

We choose a projection operator which serves to project the Liouville operator onto the relevant part of the system
dynamics, and is given by
\begin{equation}
    \mathcal{P} = \rho_l \Tr_l.
\end{equation}
Now, using \(\Tr_l \rho_l = 1\), the reduced density operator of the open central system can be
realized by the reduced density operator given by\cite{Breuer:OpenQuantumSystems}
\begin{equation}
    \rho_{S}(t) = \Tr(\mathcal{P}\rho(t)).
\end{equation}
By tracing out the variables of the reservoirs,
the Nakajima-Zwanzig equation introduces a   
quantum generalization of a master equation.\cite{Nakajima58:948,Zwanzig60:1338}
This equation is an integro-differential equation for the reduced density operator, \(\rho_S\), 
which describes the time evolution of the central system under the influence of the leads
\begin{equation}
	\partial_t \rho_S = \mathcal{P}\mathcal{L}\rho_S + \int_0^t \mathcal{K}(t,t')\rho_S(t')dt'.
\end{equation}
The kernel of the integral, \(\mathcal{K}\), describes the past history of the central system as determined by the
effects of coupling it to the environment, such as the dissipation of electrons and energy between the open central 
system and the external leads.
For our weakly coupled system, terms up to second order of the coupling to the leads, $H_T$, 
have been kept in the kernel and the Nakajima-Zwanzig equation takes the 
form\cite{Moldoveanu09:073019,Gudmundsson09:113007,Gudmundsson10:205319,Gudmundsson12:1109.4728}
\begin{equation}
        \partial_t \rho_S = \mathcal{PL}\rho_S - \sum_{l \in \{L,R\} }\Lambda[\rho]_l.
        \label{eq:NakajimaZwanzig}
\end{equation}
Here, the sum is over the leads that are coupled to the system.
The dissipative term \(\Lambda\) can be written as
\begin{equation}
      \Lambda[\rho] =  \dfrac{1}{\hbar^2} \int d\mathbf{q} \chi(t) \left\lbrace \left[\tau,\Omega[\rho]\right] + \mathrm{h.c.} \right\rbrace,
\label{eq:Dissipator}
\end{equation}
where the operator \(\tau\) is the coupling tensor of the states in the central
system and the leads.\cite{Gudmundsson09:113007}

By calculating the density of states in the leads \(D(\epsilon) = |{d\mathbf{q}}/{d\epsilon}|\)
we can represent the dissipative term from
Eq.\ (\ref{eq:Dissipator}) as
\begin{equation}
    \Lambda[\rho]= \dfrac{1}{\hbar^2}\int d\epsilon D(\epsilon)\chi(t) \left\lbrace \left[ \tau, \Omega[\rho]
        \right] + \mathrm{h.c.} \right\rbrace.
\end{equation}

Here, the coupling to the leads is switched on abruptly by the switching function \(\chi\), 
turning on the electron transport through the system in the photon cavity.
The operator \(\Omega\) describes the underlying elementary processes and the memory effects in the dissipation term
\begin{align}
      \Omega(t) =  \int_{0}^{t} ds \BiG{\chi}(s) U(t-s)& \left\lbrace \BiG{\tau}[][\dag] \BiG{\rho}(s) (1 - F) -
            \BiG{\rho}(s) \BiG{\tau}[][\dagger]F \right\rbrace\nonumber\\ \times &U^{\dagger}(t-s)\BiG{e}[][i(s-t)
            \epsilon_\mathbf{q}].
\label{eq:Omega}
\end{align}

From these operators we define corresponding superoperators
\begin{align} \label{eq:SuperoperatorsRS}
    \mathcal{S}[\rho] &=  S\rho ,\\
    \mathcal{R}[\rho] &= \rho R,
\end{align}
where,
\begin{align}
    R &= \pi (1 - F)\tau^\dagger
    \intertext{and}
    S &= \pi F \tau^\dagger ,
\end{align}
Here, \(F\) is the Fermi function which describes the equilibrium distribution of particles over the 
energy states in the leads
before the subsystems are coupled together. \(U(t-s) = \exp{\{-i(t-s)H_S/\hbar\}}\) is the unitary time-evolution 
operator for the closed central system.

\subsection{Markovian approximation}

In the ensuing discussion we extend our description to intermediate timescales as well as the steady state by
applying the Markov approximation via the operator \(\Omega\), containing the time integral describing memory
effects in the system.
In this Markovian approximation we observe that components of \(\Omega[\rho]\) induce a measure on relevant 
Bohr-frequencies in the Markovian picture.
The Bohr-frequencies correspond to a difference in energy between two many-body states. 
A key observation that we make is how the resulting component form can
be described in terms of operators as it allows us to consider the GME for an efficient
computation.

From Eq.\ (\ref{eq:Omega}), by letting \( s' = t-s\), we can write each component of \(\Omega\) as,
\begin{align}\label{eq:OmegaComponent}
    \BiG{\Omega}[\alpha \beta][\rho] =& 
            \dfrac{1}{\pi} \int_{0}^{t} ds'  
        \BiG{e}[][is'\left(E_\beta - E_\alpha - \epsilon\right)]\\ \nonumber
        &\left\lbrace \BiG{\mathcal{R}[\rho{(t-s')}]}[\alpha \beta][] -
        \BiG{\mathcal{S}[\rho{(t-s')}]}[\alpha \beta][] \right\rbrace .
\end{align}
The Markovian form of the equation can now be obtained by
assuming that \(\rho{(t-s')}\) is independent of the quantum fluctuations of the system with a period corresponding
to \( {\hbar}/(E_\beta - E_\alpha - \epsilon )\) during the time evolution, i.e.\ when \( s' \in \left[ 0,t \right] \).
Thus, \( \rho{(t-s')} \sim \rho{(t)} \equiv \rho \), and
\begin{equation} 
    \BiG{\Omega}[\alpha \beta][]\left[\rho\right] = \left\lbrace   \BiG{\mathcal{R}[\rho(t)]}[\alpha \beta][]- 
    \BiG{\mathcal{S}[\rho(t)]}[\alpha \beta][]
    \right\rbrace \int_{0}^{t} ds'\BiG{e}[][is'\left(E_\beta - E_\alpha - \epsilon\right)] .
\end{equation}
This assumption holds true for  \( t \gg \hbar/(E_\beta - E_\alpha - \epsilon ) \).  

The upper limit of the time integral can be approximated with \(t\rightarrow\infty\), giving
\begin{equation}
    \int_0^\infty ds' e^{is'(E_\beta - E_\alpha - \epsilon)} = \pi \delta(E_\beta - E_\alpha - \epsilon),
\end{equation}
where a vanishing principal part of the integral is not explicitly shown. 
A careful derivation has been given by Brasil et al.,\cite{DerivationofLindblad-Brasil} for example.
From this result we regard each component of \(\Omega[\rho]\) to induce a Dirac measure related to the component indices. That
is
\begin{equation} 
    \BiG{\Omega}[\alpha \beta][]\left[\rho\right] = \left\lbrace   \BiG{\mathcal{R}[\rho (t)]}[\alpha \beta][]- 
    \BiG{\mathcal{S}[\rho (t)]}[\alpha \beta][]
    \right\rbrace \BiG{\delta}[][\beta \alpha],
\end{equation}
where 
\begin{equation}
        \BiG{\delta}[][\beta \alpha] = \BiG{\delta}\left(E_\beta - E_\alpha - \epsilon\right).
\end{equation}
For any interaction taking place between the central system and the electron reservoirs (leads) 
as given by the coupling operator \(\tau\), 
describing the appearance or disappearance of particles from the open central system, we are only concerned 
with resonant tunneling processes.  
In this way, the Dirac measures in the Markovian form can be realized as a constraint to allowable
transitions in Liouville space.
This yields the Markovian form of the matrix elements of the dissipator from Eq.\ (\ref{eq:Dissipator})
\begin{align}\label{eq:componentdescription}
    &\Lambda_{\alpha, \beta}[\rho] \nonumber\\
    &= \dfrac{1}{\hbar^2}D^{\beta \lambda}\int d\delta^{\beta \lambda} 
        \tau_{\alpha \lambda}\left(\mathcal{R}_{\lambda \beta}[\rho (t)] - \mathcal{S}_{\lambda \beta}[\rho (t)]\right) +
        \mathrm{h.c.}\nonumber\\ 
        &-D^{\lambda \alpha}\int d\delta^{\lambda \alpha} \left(\mathcal{R}_{\alpha \lambda}[\rho (t)] - \mathcal{S}_{\alpha
                \lambda}[\rho (t)]\right)\tau_{\lambda \beta} + \mathrm{h.c.}
\end{align}

We now proceed to analyze the resulting form of the GME in a Liouville space
by seeking a description of the relation between the physical operators corresponding to the
components in Eq.\ (\ref{eq:componentdescription}).
We start by defining a matrix of Dirac measures
\begin{equation}
    \Delta_{\alpha \beta} = \delta^{\alpha \beta}.
\end{equation}
We refer to this matrix as the matrix of Dirac measures. 
It can be seen that carrying out the measures in the transposed matrix in an elementary wise fashion gives the
correspondence of indices observed in Eq.\ (\ref{eq:componentdescription}), resulting in the matrix form
\begin{equation}
    \Lambda[\rho] = \chi \int D \left[\tau, \left(\mathcal{R}[\rho] - \mathcal{S}[\rho]\right) \odot d\Delta^T \right] ,
\end{equation}
where $\odot$ stands for the elementary-wise product of matrices (or the Hadamard product). 
We can thus represent the generalized master equation as
\begin{align}\label{eq:AlgebraicNZ}
    \partial_t \rho = \mathcal{L}\rho &+  \int \chi D \left[\tau,\left(\mathcal{R}[\rho] - \mathcal{S}[\rho]\right) \odot
        d\Delta^T \right]\nonumber \\
    &+ \int \chi D \left[\left(R^\dagger\rho - \rho S^\dagger\right)\odot d\Delta, \tau^\dagger
    \right] .
\end{align}
This equation can be shown to be in a Lindblad form.
Here we have explicitly written the Hermitian conjugate term.
To solve a Sylvester equation containing this many factors we opt to use the Kronecker tensor product 
and the vectorization operator to
construct a Liouville space in which Eq.\ (\ref{eq:AlgebraicNZ}) turns into a normal linear first order
differential equation, opening the door for an efficient calculation of the Markovian form.

% Description:
% Here we cast the operators into the Liouville Tensor Space

\subsection{Markovian Time-Evolution in Liouville Space}
From Eq.\ (\ref{eq:AlgebraicNZ}) it can be seen that the time evolution of the system is given by a dissipative evolution
equation for the density matrix of the form
\begin{equation}\label{eq:evolutionfunctional}
    \partial_t \rho = \mathcal{L}[\rho] - \Lambda[\rho] .
\end{equation}
The semigroup generated by \(\mathcal{L}\) and \(\Lambda\) is given by \(e^{\mathfrak{L}t}\), where \(\mathfrak{L}\) is
a sparse matrix representation formed by the Kronecker tensor product.
The vectorspace which the semigroup acts on is the Liouville space, \(\mathbb{L}\).

This construction has been used by 
many authors.\cite{DerivationofLindblad-Brasil,Weidlich71:325,PhysRevE.91.013307,Yuge-Sugita,Petrosky01032010}

For a brief explanation of this construction we let \(\mathcal{A}\) and \(\mathcal{B}\) be operators in Fock space and
consider their sparse matrix representation as they act on the density operator \(\eta\) in an $N$-truncated Liouville
space.

The vectorization of a matrix is performed by stacking its columns underneath one another, resulting in an \(N^2\)
dimensional column vector.\cite{Henderson-1981}
That is for \(\eta = \left[\eta_1 \eta_2 \dots \eta_N\right]\), its vectorization is given by
\begin{equation} 
\vctrz{(\eta)} = \begin{bmatrix}
    \eta_1 \\
    \eta_2 \\
            \vdots \\
    \eta_N \\
    \end{bmatrix} .
\end{equation}
The operation simply expresses the isomorphism of the construction of an $N$-truncated Liouville space in terms of
    $N$-truncated Fock spaces, as expressed by \( \mathbb{L} = \mathbb{F}^{N\times N} \simeq \mathbb{F}^{N^2} \).
It is related to the Kronecker tensor product, \(\otimes\), by the identity
\begin{equation}\label{eq:BasicVctrz}
      \vctrz(\mathcal{A}\eta\mathcal{B}) = \left(\mathcal{B}^T \otimes \mathcal{A}\right)\vctrz(\eta),
\end{equation}
where
\begin{equation}
    \mathcal{B} \otimes \mathcal{A} = \left\lbrace \mathcal{B}_{\alpha \beta}\mathcal{A} \right\rbrace .
\end{equation}
In this way, the vectorization operator allows one to view the generator of the GME as a linear operator on density
operators in Liouville space. 
This is demonstrated by the vectorization of a commutator
\begin{equation}\label{eq:CommutatorVectorization}
    \vctrz{(\left[\mathcal{A},\eta\right])} = \left(\mathcal{I} \otimes \mathcal{A}  - \mathcal{A}^T \otimes \mathcal{I}
    \right) \vctrz{(\eta )} .
\end{equation}
That is, the vectorization yields a linear operator on \(\eta\) given by a linear combination of
Kronecker tensor products of Fock-operators, i.e.
\(\mathcal{I} \otimes \mathcal{A} - \mathcal{A}^T \otimes \mathcal{I} \).

In this work, we employ this for the Markovian form of the Nakajima-Zwanzig equation in terms of the reduced density operator 
for the open central system.
For an explanation of why this is useful for computations, one can consider the Liovuille-von Neumann equation for the unitary
evolution generated by \(\mathcal{A}\),
\begin{equation}\label{eq:vectorizationofL-vN}
 \partial_t \eta_{vec} = -\dfrac{i}{\hbar} \left(\mathcal{I} \otimes \mathcal{A} - \mathcal{A} \otimes
     \mathcal{I}\right)\eta_{vec}.
\end{equation}
For an $N$-truncated Fock space, the vectorization of the Liouville-von Neumann
equation results in a set of \(N^2\) linear equations with \(N^2\) components.
With modern resources of memory and the effective parallelization of matrix multiplication
for independent terms obtaining this form for Eq.\ (\ref{eq:AlgebraicNZ}) makes the Markovian form of the Nakajima-Zwanzig equation 
viable for efficient computations.

Inspired by this, we seek such a form not only for the Liouville term of the GME but also for the dissipative factors.
Thus, the necessity for describing the dissipative factors in terms of the matrices in the $N$-truncated
    Liouville space should become clear.

To treat the Hadamard product of the matrix of Dirac measures we introduce the \(\Diag\) operator which functions similarly to the
vectorization operator, but instead of stacking up columns to form a vector it forms a diagonal matrix 
\begin{equation}
 \Diag{(d\Delta )} = \begin{bmatrix}
     d\Delta_1 &  0 & \dots & 0 \\
     0 & d\Delta_2 & \dots& 0\\
     \vdots & & \ddots & \vdots\\
    0 & \dots & 0 & d\Delta_N\\ 
 \end{bmatrix} .
\end{equation}
In this presentation, column vector \(\alpha\) acts as a constraint on transitions from each many-body state in the Fock
space to the state \(|\alpha)\). 
The vectorization of a factor involving the matrix of Dirac measures thus becomes
\begin{equation}\label{eq:DeltaVctrz}
    \vctrz\left(\int X \odot d\Delta\right) = \int \Diag\left(d\Delta\right)\vctrz(X),
\end{equation}
from which follows the vectorization of the dissipative terms from Eq.\ (\ref{eq:evolutionfunctional}),
\begin{align}  
    &\vctrz\left( \Lambda[\rho] \right) =
      \vctrz{\left( \int D \left[\tau,\left\lbrace \mathcal{R}[\rho] -
              \mathcal{S}[\rho]\right \rbrace \odot d\Delta^T\right]\right)} \nonumber\\
        &=\int \left(I \otimes D\tau - D\tau \otimes I\right)\vctrz{\left( \left\lbrace \rho R -
            S\rho\right\rbrace \odot d\Delta^T\right)},
\end{align}
where we have used Eq.\ (\ref{eq:CommutatorVectorization}) and (\ref{eq:SuperoperatorsRS}).
Now for convenience in constructing the terms programmatically, we opt for seperating the integral by linearity
\begin{align}\label{eq:DiracVCTRZ}
    \vctrz\left( \Lambda[\rho] \right) &= \int \left(I \otimes D\tau - D\tau \otimes I\right)\Diag{(d\Delta^T)}\\\nonumber
     &\times \int\Diag{(d\Delta^T)} \left(R^T \otimes I - I\otimes S
        \right)\vctrz{(\rho)}.
\end{align}
Notice that if \(X\) is a non-symmetric operator in Fock-space we have,
\begin{equation}
    \int \Diag{(d\Delta)}X \neq \int X \Diag{(d\Delta)},
\end{equation}%
as in the left case we are measuring the $i$-th line of \(X\) by the $i$-th measure but in the right case we have the $i$-th
column of \(X\) by the $i$-th measure. 

The sequence of linear operators which we have now obtained can be written in the form
\begin{equation}
 \Lambda_{vec} =  \sum_{i=1}^4\mathfrak{Z}_{i 1}\mathfrak{Z}_{i 2},
\end{equation}
where we have 
\begin{align}\label{eq:Zi}
 \mathfrak{Z}_{i 1} &=  (-1)^{i+1} \int(I \otimes D\tau) \Diag{(d\Delta^T)}\qquad \text{for } i = 1,2\\
 \mathfrak{Z}_{i 1} &= (-1)^{i+1}\int ( D\tau \otimes I) \Diag{(d\Delta^T)}\qquad \text{for } i = 3,4\\
 \mathfrak{Z}_{i 2} &= \int \Diag{(d\Delta^T)}(R^T \otimes I) \qquad \text{for } i = 1,3\\
 \mathfrak{Z}_{i 2} &= \int \Diag{(d\Delta^T)}(I \otimes S) \qquad \text{for } i = 2,4.
\end{align}
In our previous publication a general element from this sequence is described to present the method to the
same effect.\cite{2016arXiv160508248G} 
It is worthy to note that in this discussion identifying the appearance of the Dirac delta functional from the Markov
approximation in terms of a measure is beneficial for our notation apart from offering a clear algebraic convenience. 
Furthermore, it makes the splitting up of terms easy
to understand in terms of linearity as we have done in Eq.\ (\ref{eq:DiracVCTRZ}). 
We see clearly from this that applying a Markovian approximation to a
non-Markovian representation of the dynamics of the system is done by measuring the most effective transitions between
many-body states which give rise to the steady state. 
The Hermitian conjugate terms indicated in Eq.\ (\ref{eq:componentdescription}) are given by
\begin{equation}
    \vctrz{(\mathrm{h.c.})} = \vctrz{\left( \int D \left[\left(R^\dagger\rho - \rho S^\dagger\right)\odot d\Delta,
                \tau^\dagger\right]\right)}.
\end{equation}
We treat it similarly and end up with,
\begin{align}\label{eq:HCDiracVCTRZ}
     \vctrz{(\mathrm{h.c.})} = \int\left(D\tau^\dagger \otimes I - I \otimes D\tau^\dagger\right) \Diag{(d\Delta)} \\\nonumber
        \times\int \Diag{(d\Delta)} \left(I \otimes R^\dagger - S^C \otimes I\right) \vctrz{(\rho)} .
\end{align}
That is, in contrast to before we now obtain
\begin{equation}
 (\mathrm{h.c.})_{vec} =  \sum_{i=1}^4\mathfrak{X}_{i 1}\mathfrak{X}_{i 2},
\end{equation}
where,
\begin{align}\label{eq:Xi}
 \mathfrak{X}_{i 1} &= (-1)^{i+1} \int (D\tau^\dagger \otimes I)\Diag{(d\Delta)} \qquad \text{for } i = 1,2\\
 \mathfrak{X}_{i 1} &= (-1)^{i+1}\int ( I \otimes D\tau^\dagger)\Diag{(d\Delta)} \qquad \text{for } i = 3,4\\
 \mathfrak{X}_{i 2} &= \int  \Diag{(d\Delta)}(I \otimes R^\dagger)\qquad \text{for } i = 1,3\\
 \mathfrak{X}_{i 2} &= \int \Diag{(d\Delta)}(S^C \otimes I) \qquad \text{for } i = 2,4.
\end{align}

In previous work,\cite{2016arXiv160508248G} we have encapsulated all the terms contained in the
dissipator by ways of treating the general form of the factors in the sequence
\begin{equation}
    Z = \int DA\left\lbrace(\mathcal{R}[\rho] - \mathcal{S}[\rho])\odot d\Delta^T\right\rbrace B
\end{equation}
with \(A\) and \(B\) as placeholders for any operators in the Fock space. 
The Hermitian conjugate term can then be given by
\begin{equation}
    Z^\dagger = \int DB^\dagger\left\lbrace(\mathcal{R^\dagger}[\rho] - \mathcal{S^\dagger}[\rho]) \odot d\Delta
    \right\rbrace A^\dagger .
\end{equation}
We can thus rewrite Eq.\ (\ref{eq:evolutionfunctional}) in the Liouville tensor space as a linear first order differential
equation in Liouville space
\begin{equation}\label{eq:Liouvilleevolutionfunctional}
    \partial_t \rho_{vec} = -\dfrac{i}{\hbar}\left(I \otimes H - H \otimes I + \Lambda_{vec} \right)\rho_{vec}.
\end{equation}
In order to compute and solve this equation we need to compile the matrices representing the operators in the truncated Liouville
space. For a fixed number of timesteps we calculate the dynamical maps given by the Markovian semigroup. 

We shall describe how the generator of the semigroup is compiled computationally in the appendix. 

We write the generator of the Markovian semigroup as
\begin{equation}
\mathfrak{L} = -\dfrac{i}{\hbar}\left(I \otimes H - H \otimes I +
        \Lambda_{vec}\right),
\end{equation}
thereby allowing us to write Eq.\ (\ref{eq:Liouvilleevolutionfunctional}) compactly as,
\begin{equation}\label{eq:finalpde}
    \partial_t \vctrz(\rho) = \mathfrak{L} \vctrz(\rho).
\end{equation}
This equation can be solved exactly in Liouville space by performing matrix diagonalization on \(\mathfrak{L}\). 
As $\mathfrak{L}$ is not Hermitian this is done by finding the
left and the right eigenvectors, \(\mathfrak{U}\) and \(\mathfrak{B}\), satisfying, 
\begin{align}
    \mathfrak{LB} &= \mathfrak{B L}_\text{diag},\\
    \intertext{and}
    \mathfrak{UL} &= \mathfrak{L}_\text{diag}\mathfrak{U},\\
    \intertext{with}
    \mathfrak{UB} &= I\\
    \mathfrak{BU} &= I.
\end{align}
The diagonal matrix is \(\mathfrak{L}_\text{diag}\) and \(I\) is the identity operator.
Eq.\ (\ref{eq:finalpde}) can now be solved as
\begin{equation}\label{eq:solution}
    \vctrz(\rho(t)) =
    \left\{\mathfrak{U}\left[\exp{\left(\mathfrak{L}_\text{diag}t\right)}\right]\mathfrak{B}\right\}\vctrz(\rho(0)).
\end{equation}
As the time evolves the collection of the exponential terms in the component equation, which we investigate in Section
2.1, approach the Dirac measure. Thus, for our system, the limit of this equation is a reliant way to seek the steady
state. As a note of caution we like to mention that subroutines in Lapack and derivatives thereof for finding the 
eigenvalues and vectors of a general complex non-Hermitian matrix do use another convention for the normalization
of the left and right eigenvectors.

%--------------------------------------------
\section{Computational Results}
The computational results presented below have been obtained for a system requiring 120 many-body states
in the Fock-space ${\mathbb{F}}$. As we are concentrating on the long time behavior of the system we exponentially
distribute 84 time points from 1 ps to 1 s, thus reducing our attention on transient behavior that might
still be present in the initial time regime. There are two main reasons for our quest to write the 
Markovian time evolution of the master equation in terms of the matrices we have introduced. The first one
is to gain knowledge about the fundamental processes inherent in the system, and the second one is to
use modern parallelization techniques that have been perfected with linear algebra subroutines for FORTRAN
or C. We use Intel FORTRAN MKL BLAS and Lapack routines for all linear algebra tasks on CPU-clusters, and 
add CUDA, CUBLAS and MAGMA on mixed CPU-GPU-clusters.  

We select four different eigenstates of the closed electron-photon system as initial states for the transport
calculations. The bias window is fixed by setting $\mu_L=1.4$ meV, and $\mu_R=1.1$ just above the band bottom
of the leads at 1.0 meV, and the plunger gate voltage that varies the electronic energy levels of the central system 
with respect to the leads is set at $V_g=-1.9$ mV in order to have several states below the bias window.  
The four initial states are the vacuum state, the state with only one photon, and the one with only two photons, but
no electrons. In addition, we select the lowest one-electron state with a nonzero photon component. This last state 
happens to be a Rabi-split replica of the one-electron ground state with the non-integer photon mean value of 0.8. 
These initial states are chosen as to demonstrate the order of magnitude in time scales that we can observe. 
Previously, very long relaxation times of the system have prevented us from using the non-Markovian approach to clearly
observe the time scales needed for determining the steady-state of the system. 
We use GaAs parameters, $m^*=0.067m_e$, and $g=-0.44$, and set the coupling to the leads at $g_{\mathrm{LR}}a^{3/2}_w=0.124$
meV. $B=0.1$ T, and in the leads before coupling the temperature $T=0.5$ K. 

\begin{figure}[htb]
    \centering
    \includegraphics[width=0.48\textwidth]{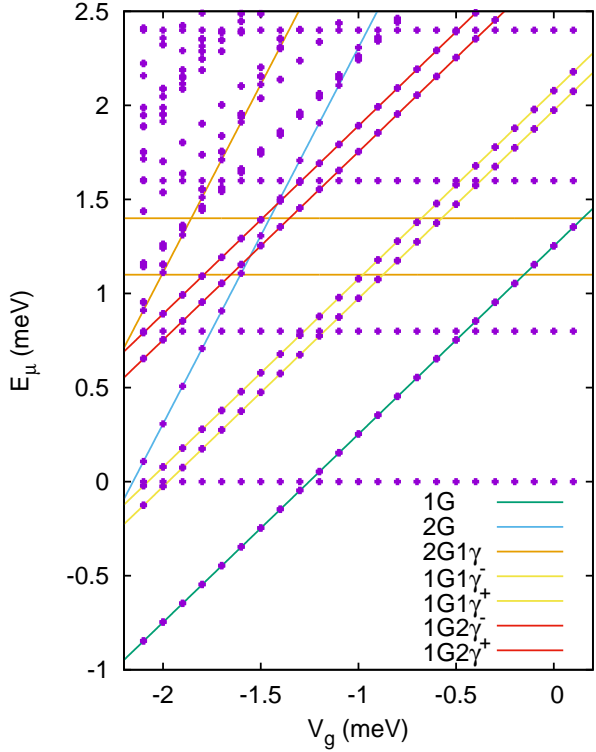}
      \caption{The many-body energy spectrum of the closed central system as
               a function of the plunger gate voltage $V_g$. The golden horizontal lines indicate the
               chemical potential of the left lead $\mu_L=1.4$ meV, and the right lead $\mu_R=1.1$ meV.
               1G denotes the one-electron groundstate, 2G the two-electron one, and 1G1$\gamma^\pm$
               stands for the Rabi-split first photon replica of the 1G. $B=0.1$ T, $g_{\mathrm{EM}}=0.05$ meV,
               $m^*=0.067m_e$, and $g=-0.44$. The horizontal purple dotted lines are photon states with zero electrons. 
            }
\label{Fig-Erof-Thj}            
\end{figure}

Before introducing the results for the open system, we look at the many-body energy spectrum of the closed system as a function of 
the plunger gate voltage, \(V_g\), in Fig.\ \ref{Fig-Erof-Thj}.
The change in energy of each many-body state with respect to the plunger gate voltage depends linearly on the number of
electrons in the state explaining a steeper slope of energy for states containing two electrons. 
We observe the occurance of a Rabi splitting of electron states when the energy of the photon cavity is close to the Bohr frequency 
of some particular electron states. 
(A superposition of Rabi split states represents a state where the photons are repeatedly absorbed and emitted). 

\begin{figure}[htb]
    \centering
    \includegraphics[width=0.48\textwidth]{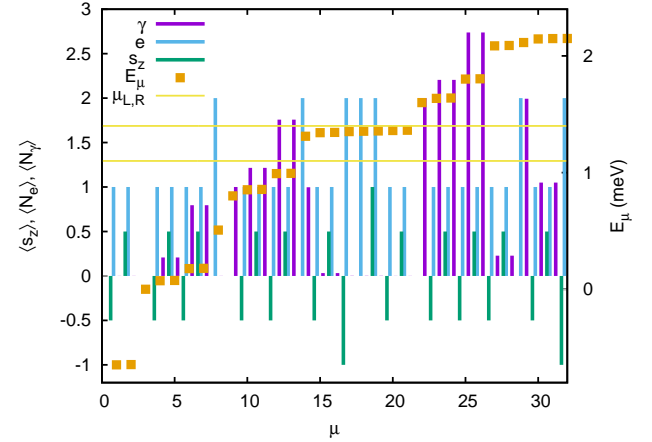}
    \caption{The energy, the electron number \(N_e\), the mean photon number \(N_\gamma\), and the spin \(s_z\) for each
        many-body state \(|\mu)\) for \(V_g = -1.9\) meV. The yellow horizontal lines indicate the bias window. Other
        parameters as in Fig.\ \ref{Fig-Erof-Thj}}
    \label{Fig-NeNphESz}
\end{figure}

A more detailed picture of the properties of the many-body states close to the bias window can be seen in Fig.\ \ref{Fig-NeNphESz}
for the case of \(V_g = -1.9\) mV.
In it we observe, for the lowest dressed many-body states \(|\mu)\), the energy as well as the photon component and
the electron component along with their spin components.
For the time evolution we will observe how these states are charged or occupied according to the Markovian time
evolution, before, ultimately converging to the the steady state. 
The Rabi splitting seen in Fig.\ \ref{Fig-Erof-Thj} is reflected in non-integer photon mean number for 
many states in Fig.\ \ref{Fig-NeNphESz}. The close to resonance situation between the photons and the electrons in
the system seen by Rabi splitting makes the perturbational picture of photon replicas of electron states not
adequate for the photon-dressed electron states we have. 
The vacuum state is $|3)$ with zero energy as the plunger gate
voltage -1.9 mV pulls the two spin components of the one-electron ground state, $|1)$ and $|2)$, to 
negative energy values. The two-electron ground state is $|8)$.

\begin{figure}[htb]
      \centering
      \includegraphics[width=0.45\textwidth]{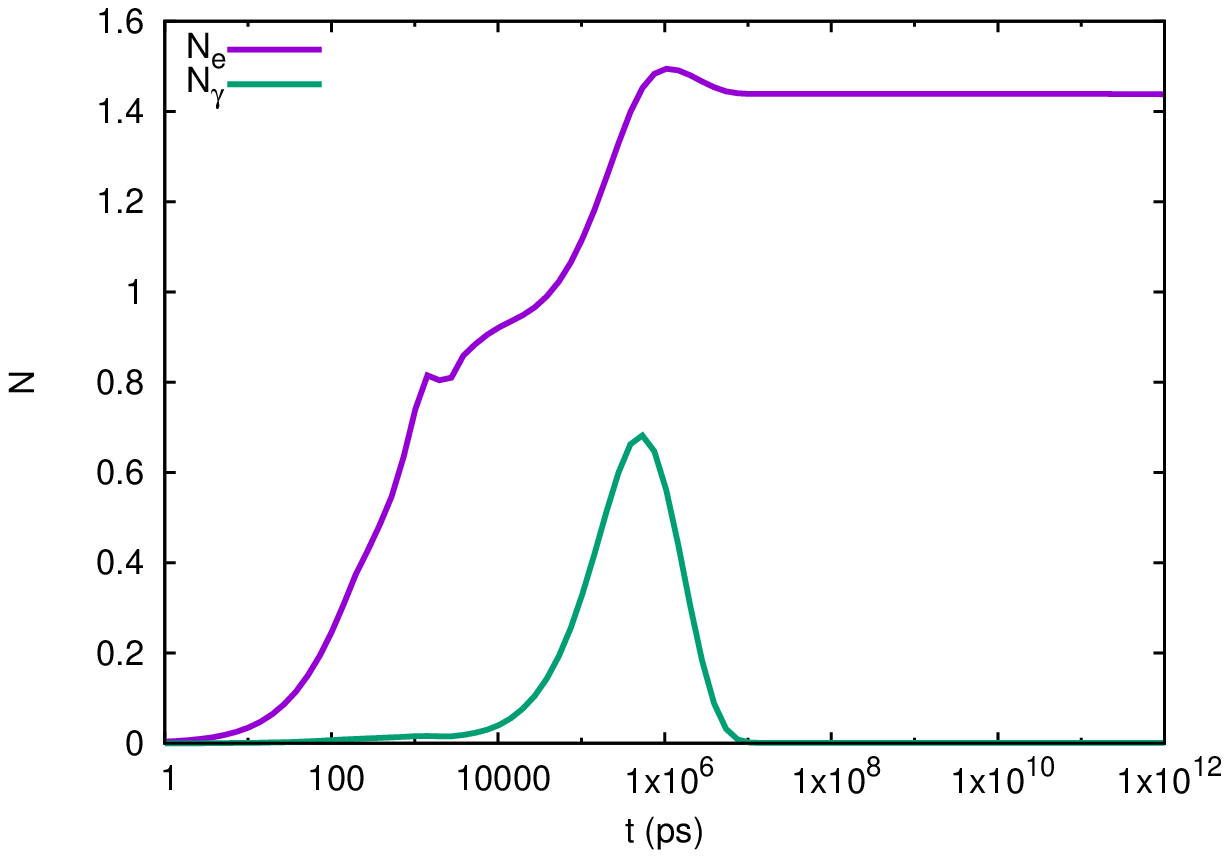}
      \includegraphics[width=0.45\textwidth]{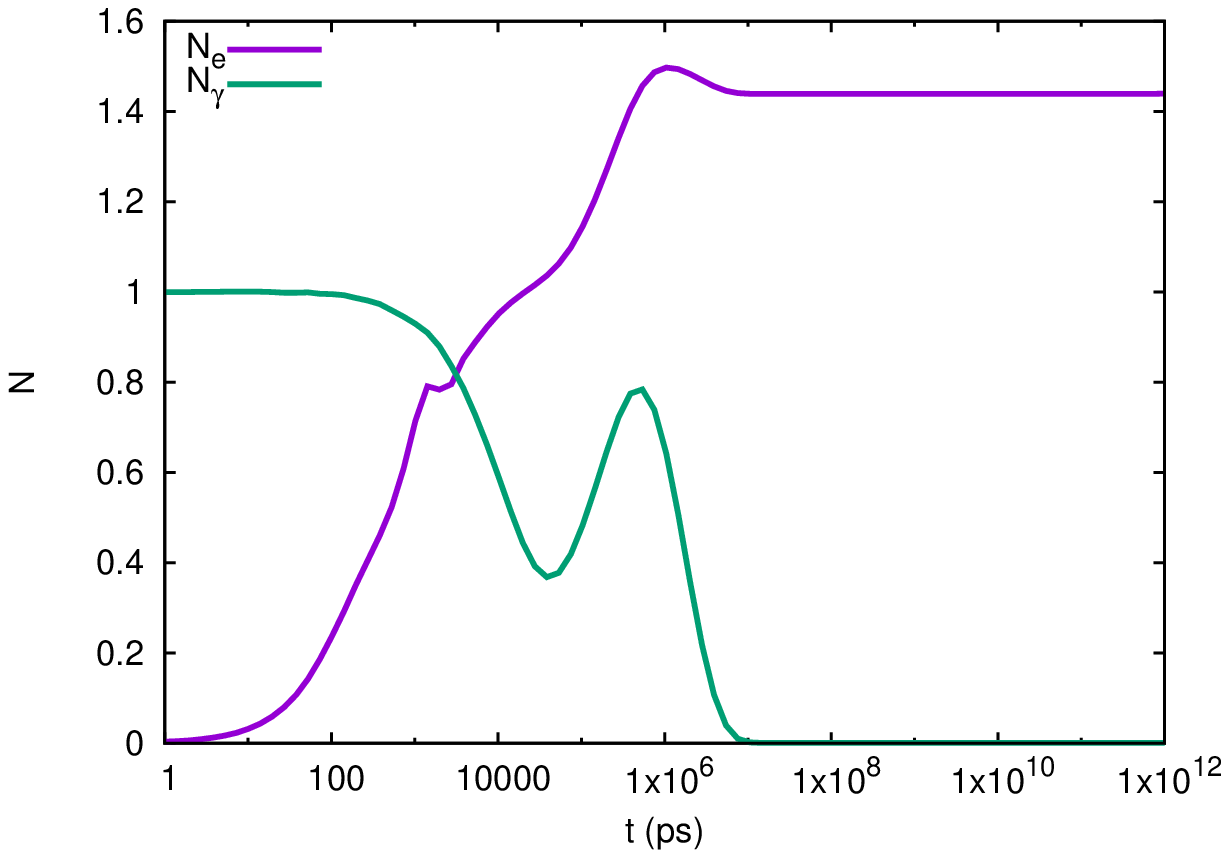}
      \includegraphics[width=0.45\textwidth]{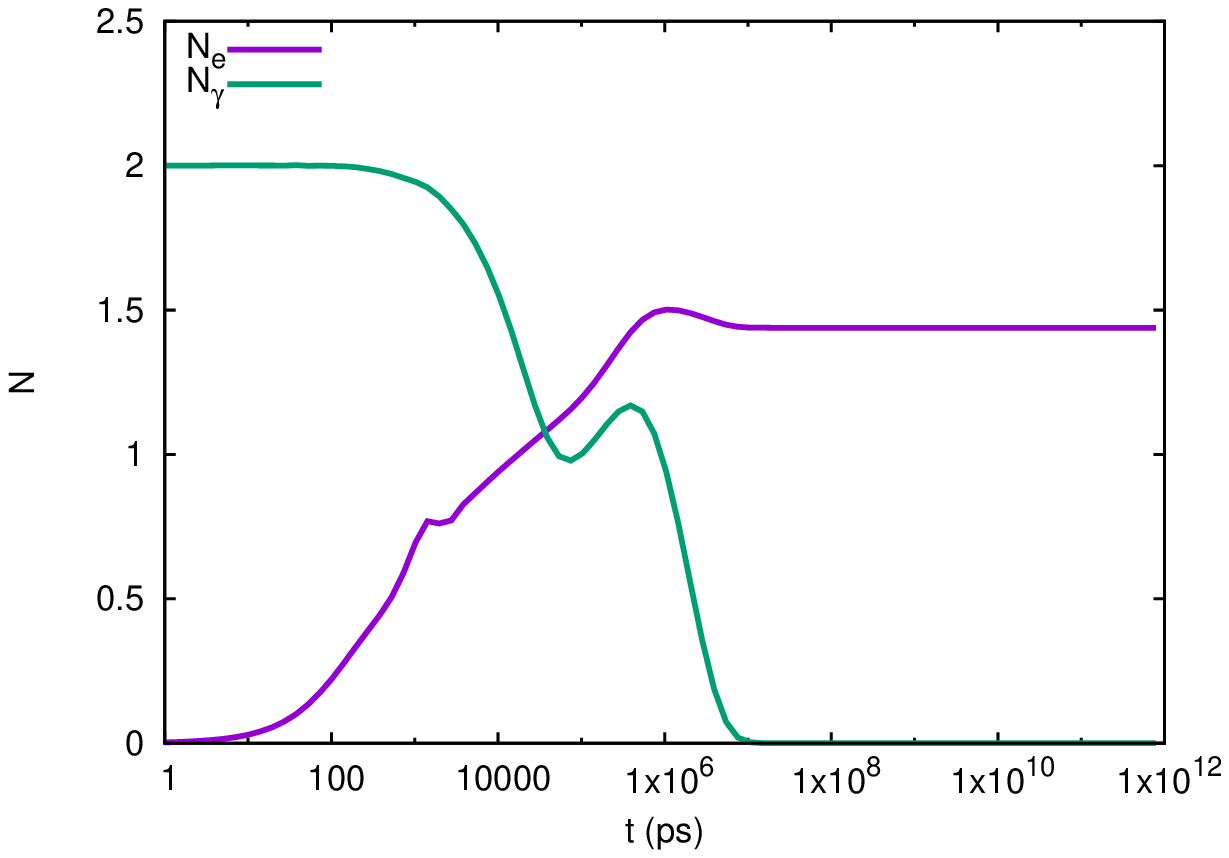}
      \caption{The mean values of the total electron (purple) and photon (green) numbers
               as functions of time for $V_g=-1.9$ mV. The initial number of photons in the cavity is zero (top panel),
               one (center panel) or two (bottom panel). The initial state is the vacuum state $|3)$.}
\label{Fig-NeNg}
\end{figure}

The Markovian time evolution of the mean electron and photon numbers for the system from the initial vacuum state, $|3)$, 
is presented in the upper panel of Fig.\ \ref{Fig-NeNg} for \(V_g=-1.9\) mV.
The Markovian time-evolution is expressed according to the solution of Eq.\ (\ref{eq:solution}) and is shown 
in Fig.\ \ref{Fig-NeNg} on a logarithmic time scale.
Looking at the transient regime we see that a non-zero mean number of photons 
appears in the time interval \(0.1 - 100\) \(\mu\)s.
The center panel of Fig.\ \ref{Fig-NeNg} shows the evolution of the mean number of electrons and photons
when one photon is initially in the system, i.e.\ the inital state is $|9)$,
and the bottom panel displays the evolution of the initial state with two photons, $|22)$.
For all three cases the system has reached the same steady state for time $t>10^{-4}$ s.

\begin{figure}[htb]
      \centering
      \includegraphics[width=0.45\textwidth]{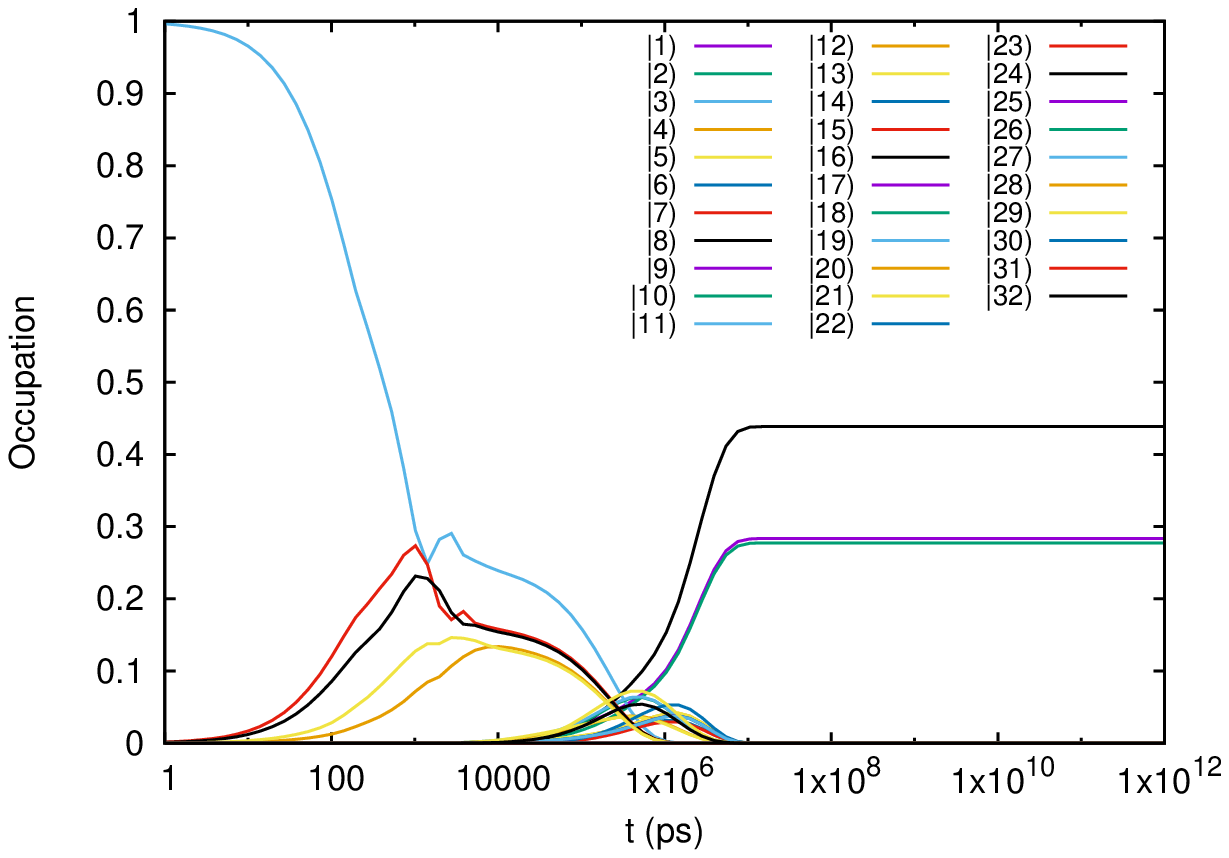}
      \includegraphics[width=0.45\textwidth]{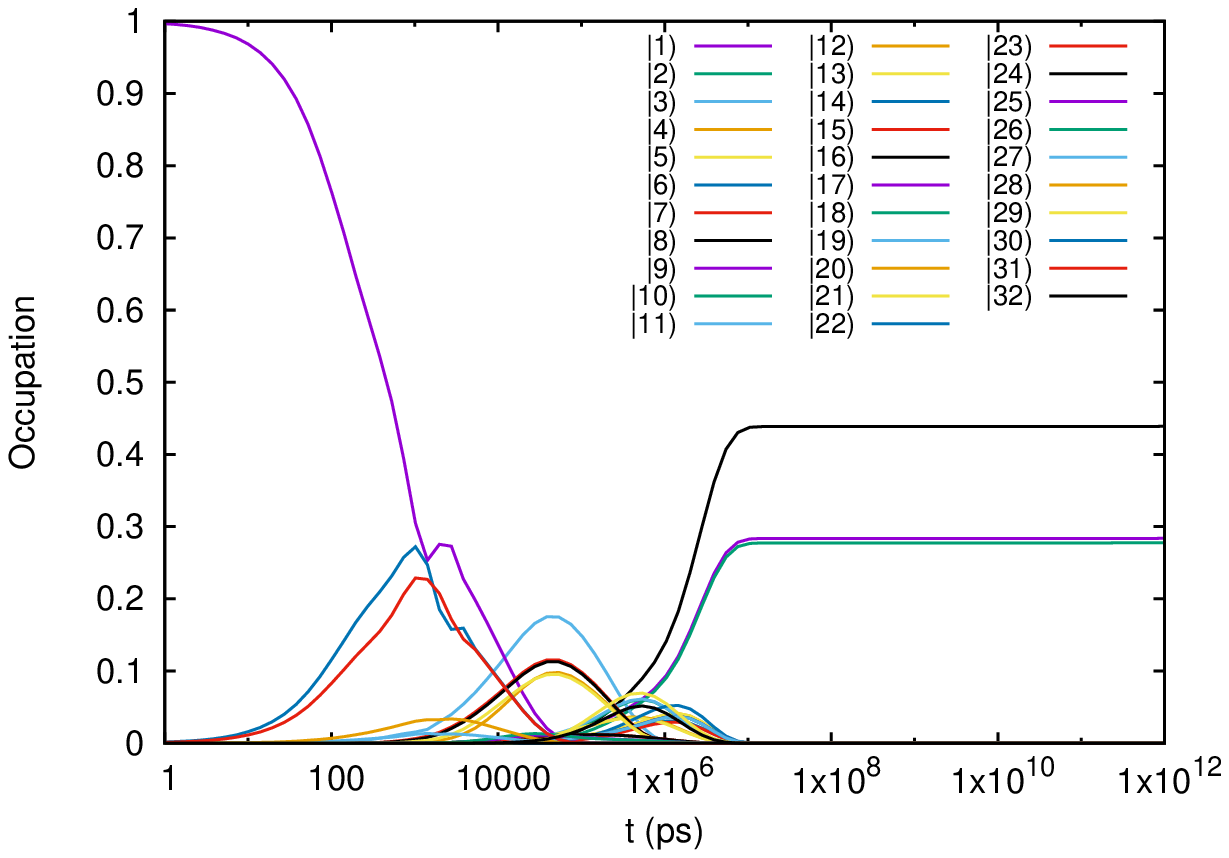}
      \includegraphics[width=0.45\textwidth]{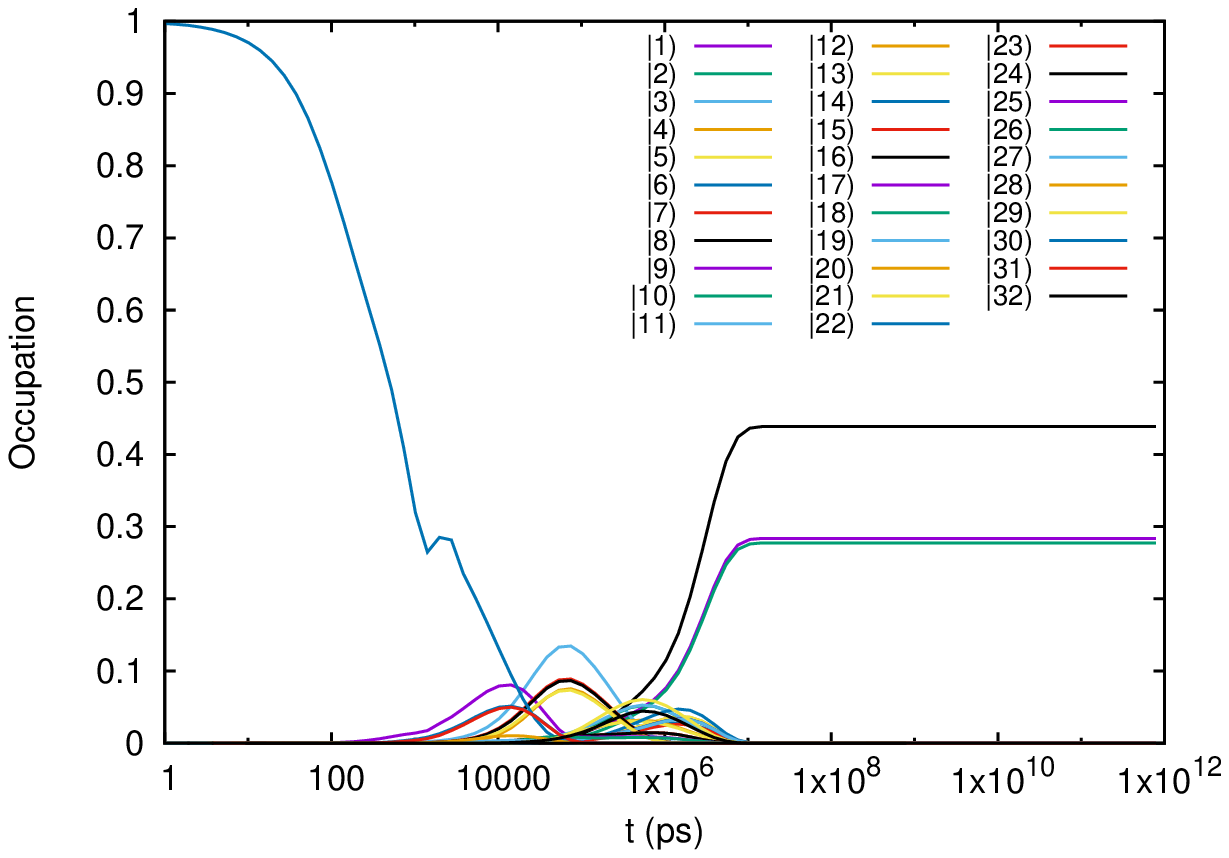}
      \caption{The mean occupation of the many-body states as a function of time, for $V_g=-1.9$ meV.
               Only states with relevant occupation are listed.
               The initial number of photons in the cavity is zero (top panel),
           one (center panel) or two (bottom panel).}
       \label{Fig-occ}
\end{figure}

By viewing the occupation of the many-body states during the time evolution as in Fig.\ \ref{Fig-occ}
we can see which intermediate state participate in the transport, or in other words here, 
which states serve as intermediate states for the system on its path to the steady state.
These figures clearly show how the different transient history for each of the examples converges to the same steady
state in different manners, and that the steady state is a mainly comprised of contibutions from the 
the two-electron ground state, $|8)$, a spin singlet, and both spin components of 
the one-electron ground state, $|1)$ and $|2)$. 

\begin{figure}[htb]
      \centering
      \includegraphics[width=0.45\textwidth]{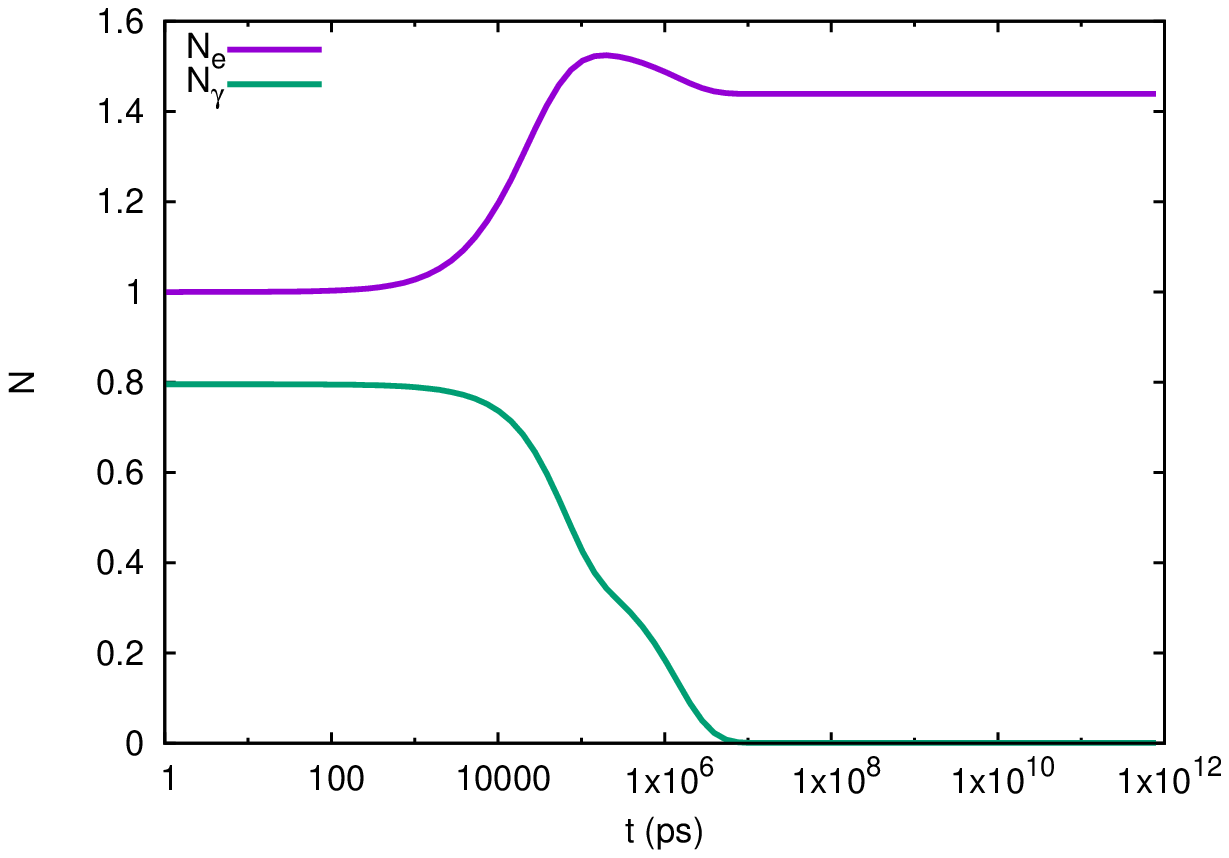}
      \includegraphics[width=0.45\textwidth]{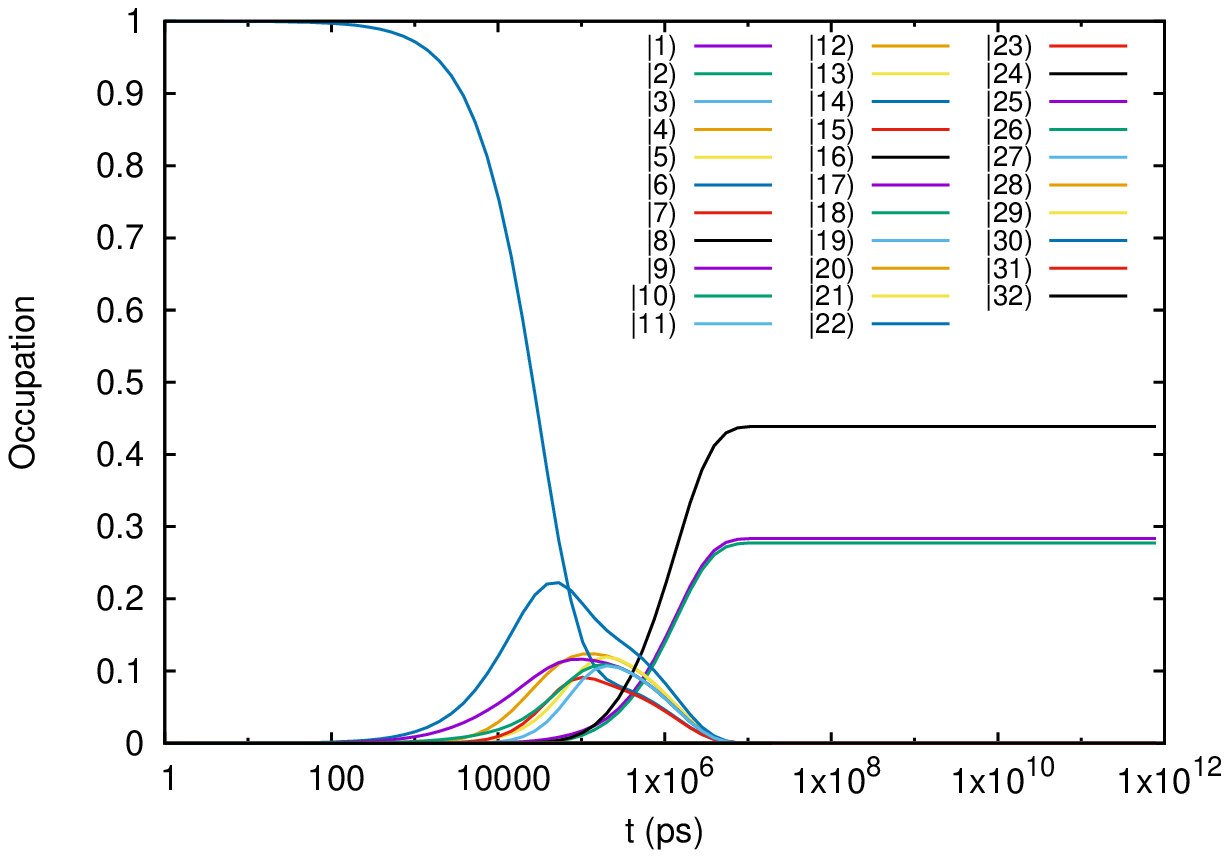}
      \caption{(Top panel), the mean values of the total electron (purple) and photon (green) numbers
               as functions of time.  (Lower panel) The mean occupation of the many-body states as a function of time.
               $V_g=-1.9$ mV. The initial state is the Rabi-split lowest photon replica of the single-electron
               ground state $|6)$.}
  \label{Fig-NeNg2}
\end{figure}

For all three scenarios the steady state is mainly composed of the states \(|1), |2)\) and \(|8)\), where \(|1)\) and
\(|2)\) are the two spin components of the one-electron ground state.
State \(|8)\) is the
two-electron ground state, \(|9)\) is the one-photon state, and \(|10)\) is a one-electron state
with approximately \(0.02\) 0-photon, \(0.75\) 1-photon and \(0.23\) 2-photon contribution.
With zero photons in the cavity at time \(t=0\), we get a considerable charging of state \(|15), |16), |20)\) and \(|21)\), which are
one-electron states, just above the bias window in Fig.\ \ref{Fig-NeNg}. 
With a single photon in the cavity at the beginning we get a charging of states \(|30)\) and
\(|31)\). These are one-electron-one photon states with spin up and down respectively which are in the upmost
right corner above the bias window in Fig.\ \ref{Fig-NeNg}. 
In both cases the charging starts at around \(t \sim \vspace{0.1cm} 10\) ps and lasts until \(t \sim 1000 \vspace{0.1cm}\) ps 
when occupation of the vacuum state and one-electron states with many photons starts. 
The occupation of the vacuum state in all three scenarios occurs at
\(t = 100\) \(\mu\)s to \(t=1 \vspace{0.1cm}\) \(\mu\)s, preceding a charging of mixed states before the steady
state is reached before \(t > 10\) \(\mu\)s. 
With two photons the steady-state remains the same, but since states with a two-photon component are situated high above
the bias window we do not see a charging of those states in the cavity.

\begin{figure}[htb]
    \centering
    \includegraphics[width=0.8\textwidth]{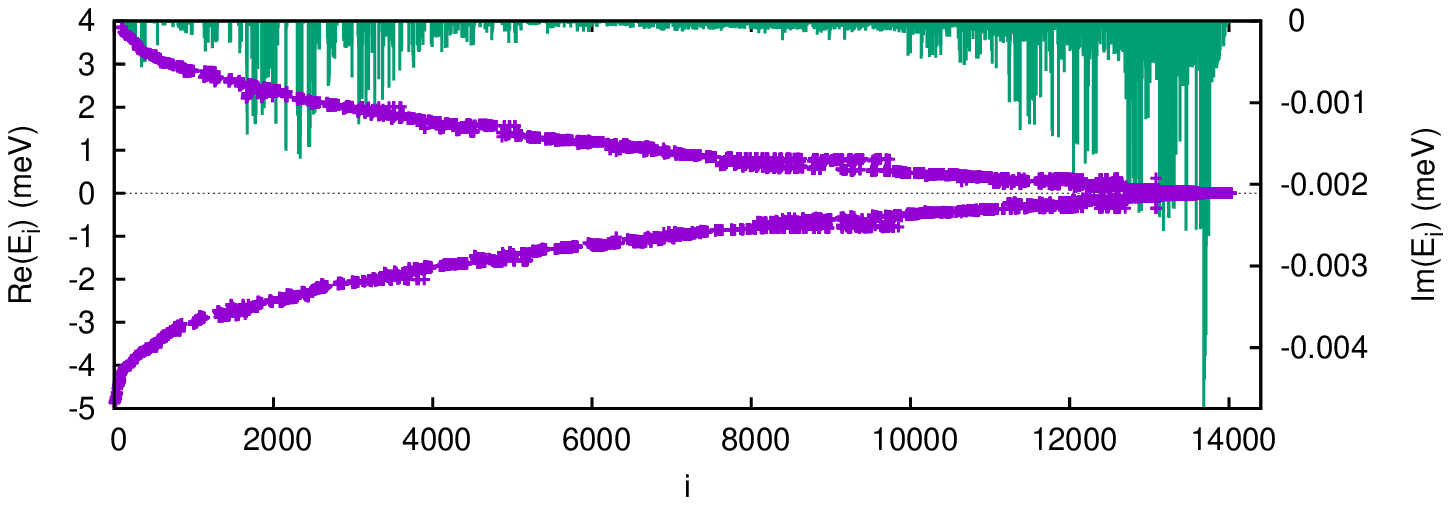}
    \includegraphics[width=0.8\textwidth]{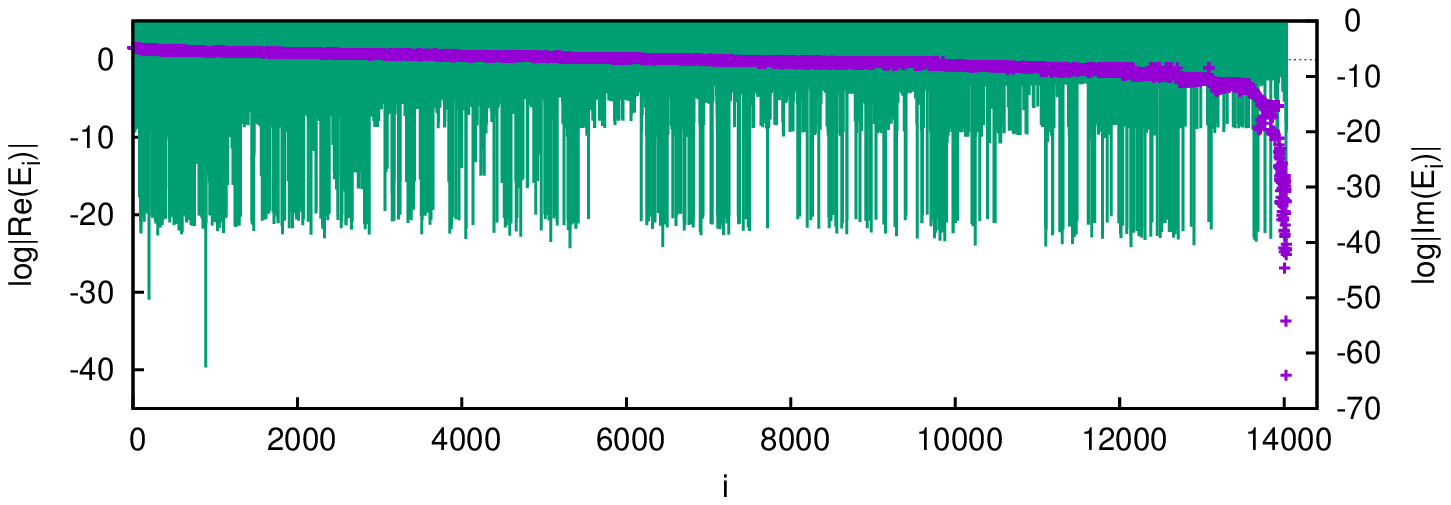}
    \caption{The complex spectrum of the Liouville operator (upper panel). The real part is indicated
             with violet dots, and the imaginary part by green impulses. (Lower panel) the 
             natural logarithm of the absolute values of the real and imaginary parts of the eigenvalues
             in order to highlight the zero eigenvalue.}
    \label{eigenspectrum_L}
\end{figure}

All the senarios presented here can be related back to the spectrum of the Liouville operator through
the solution (\ref{eq:solution}) of the evolution equation (\ref{eq:finalpde}). The spectrum of the
Liouville operator is displayed in Fig.\ \ref{eigenspectrum_L}
\onecolumngrid

\twocolumngrid
For a closed system the Liouville spectrum has only real eigenvalues reflecting all possible Bohr frequencies 
of the system. In the case of an open system we can view the spectrum as displaying all transitions and their
respective life times. With this in mind we observe that for our system the time evolution is composed of
many transitions with a large spread of life times, exactly what one would expect after viewing Figures
\ref{Fig-NeNg}-\ref{Fig-NeNg2}. 

The calculations show that any information about the transient behaviour of the system can not be retrieved
from its steady state for the selected scenarios. With photons initially in the cavity entering electrons are promoted
electromagnetically to states above the bias window. As time evolves these electrons cascade down to the steady state,
which remains the same for the different configurations we consider. 

In Figure \ref{Fig-NeNg} we notice a small step-like struture in $\langle N_e\rangle$ around $t\approx 1000$ ps. 
This structure is caused by slow relaxation of the two spin components of particlar states as can be verified 
by inspecting the time-dependent occupation of the many-body states presented in Fig.\ \ref{Fig-occ} having 
in mind their properties as shown in Fig.\ \ref{Fig-NeNphESz}.

\section{Conclusions}
In this article we present a general eficient computational framework for calculating a Markovian solution of a Nakajima-Zwanzig
equation in Liouville space for a complex open quantum system. 
This framework is diligently derived by observing how the Markov approximation can be
expressed algebraically such that an algorithmic composition becomes evident.
Using this framework we can effectively calculate the time evolution of the system at all time
scales without having to resort to a time consuming numerical integration of the equation of motion. 
We have displayed the value of this approach for an electron system weakly coupled to external leads, 
but strongly coupled to far-infrared photons of a cavity, in which the electron system is embedded. 
Both the weak coupling to the leads and the frequency of the cavity photons can lead to very long relaxation 
times that can not be ignored in the system as we place the bias window, defined by the leads, high into
the many-body spectrum of photon-dressed states of the central system. We observe a cascade of radiative
and nonradiative intermediate transitions as the system approaches the steady state.

Even though we implement the Markovian approximation and the mapping of the problem to Liouville space for a particular
model in the weak coupling to the environment, nothing in our scheme is dependent on this model. 
The Markovian solution becomes attainable by introducing a matrix of Dirac measures which selects the relevant many-body state
transitions describing the evolution of our system.

The fact that we can analyse the time evolution of an open electron system by ``exactly interacting'' few electrons
and many photons in a nontrivial geometry opens up the possibility to investigate many regimes of interesting physical
phenomena that are not easily accessible to simpler model systems.

%************************************************
%************************************************
% Computational model
% Description: Here we describe how we compile the components of the equation
%************************************************
%************************************************
%\pagebreak
\section{Appendix I - The Composition of Factors in Liouville Space}

In this Appendix, we explain how the sparse matrix representations of dissipative factors in Liouville space can be
generated computationally. 
This is accomplished by showing that the block matrix structure of these factors, which emerges from the Markovian form, 
can be derived in a clear and
an efficient manner.
It can be seen that the Dirac measures enforce a diagonal dominance in the sparse matrix structure, by eliminating either
off-diagonal block matrix components or off-block-diagonal elements.
This property can be exploited to derive a simple algorithm for the compilation of dissipative terms
in the truncated Liouville space.

First, we give a comprehensive description of how the Nakajima-Zwanzig equation can be compiled by disseminating the
total Liouvillian \(\mathfrak{L}\) into its respective terms. 
The sparse matrix representation of the generator \(\mathfrak{L}\) is obtained as a linear combination of Kronecker 
tensor products describing the unitary evolution of the system, as well as the dissipative factors given 
by Eq.\ (\ref{eq:Zi}) and (\ref{eq:Xi}).

For this, we shall describe the process explicitly in terms of the block matrix structure of the
factors of \(\mathfrak{L}\). Not only does this provide a mathematical convenience, it also reveals a simple structure
which can be explored in a computational manner.

In our notation, block matrix \((\alpha,\beta)\) of the total Liouvillian is given by \(\mathfrak{L}^{\alpha}_{\beta}\).

First, blockmatrix \( (\alpha,\beta)\) of the positive term, given by the unitary evolution for the system is expressed as,
\begin{equation}
    (I \otimes H)_{\beta}^{\alpha} =
 \begin{cases} 
              H & \qquad\text{if $\alpha = \beta$} \\
              0 & \qquad \text{otherwise} ,
 \end{cases}
\end{equation}
i.e.\ as a block diagonal matrix. 
Here, each block matrix on the diagonal represents the Hamiltonian for a given transition
in Liouville space.
The block matrices are elements in a subspace of the \(N\)-truncated Fock space. 

The negative term of the unitary evolution is expressed by
\begin{equation}
 (H^T \otimes I)^{\alpha}_{\beta} = I_N H_{\beta,\alpha},
\end{equation}
where \(I_N\) is the identity and \(H_{\beta,\alpha}\) is the \(\beta,\alpha\) component of the system Hamiltonian.
Here, each of the block matrices is an \(N\) diagonal blockmatrix containing information about the states corresponding
to its block matrix number.

The dissipative terms can also be written in this way from the sparse Markovian description given by
Eq.\ (\ref{eq:Zi}) and (\ref{eq:Xi}). Whereas, a mixed multiple of factors such as, \(\tau\rho\tau^\dagger\), would, in the
non-Markovian case, break down the `simple' sparse description given for the unitary evolution,  the diagonal action of the Dirac
matrix eliminates components which would otherwise contribute to the evolution.
Moreover, this elimination of terms can be exploited, for each of the dissipative
factors such that they can be compiled effectively with time complexity  \(O(N^3)\). 
Importantly, this compilation can be performed in parallel for all factors thus enabling very fast calculations.

The matrix of Dirac measures can be written in Liouville space in terms of its columns as diagonal block matrices by
\begin{equation}
    \Delta = \begin{bmatrix}
        \Diag{(\Delta^1)} & 0 &  \dots & 0\\
        0 &  \Diag{(\Delta^2)} &   \dots & 0\\
        \vdots & & \ddots & \vdots \\
        0 & \dots & &  \Diag{(\Delta^N)} \\
    \end{bmatrix}.
\end{equation}

Here, a diagonal Dirac blockmatrix \(\Diag{\Delta^{\alpha}}\) measures any transition to the many body state \(|\alpha)\). Acting from the
left, it applies the measure to each column of an operator, but to each row if acting from the right. For the transpose
of the Dirac delta, diagonal blockmatrix number \(\Diag{\Delta_{\alpha}}\) measures any transition from the many-body state \(|\alpha)\).  

For demonstration we shall consider the sparse representation of the two dissipative terms \(\mathfrak{Z}_{21}\mathfrak{Z}_{22}\) and
\(\mathfrak{Z}_{31}\mathfrak{Z}_{32}\) from Eq.\ (\ref{eq:Zi}) and (\ref{eq:Xi}).

In terms of their block matrix structure, these terms are written as,

\begin{align}\label{eq:BlocksZi}
 (\mathfrak{Z}_{2 1})^{\alpha}_{\beta} &= - \int(I \otimes D\tau)^{\alpha}_{\beta} \Diag{(d\Delta^{\alpha})}\\
(\mathfrak{Z}_{2 2})^{\alpha}_{\beta} &= \int \Diag{(d\Delta^{\alpha})}(I \otimes S)^{\alpha}_{\beta} \\
 (\mathfrak{Z}_{3 1})^{\alpha}_{\beta} &= \int ( D\tau \otimes I)^{\alpha}_{\beta} \Diag{(d\Delta^{\alpha})} \\
(\mathfrak{Z}_{3 2})^{\alpha}_{\beta} &= \int \Diag{(d\Delta^{\alpha})}(R^T \otimes I)^{\alpha}_{\beta} \\
\end{align}
By referring to the block matrix description in this way, one can identify the
component structure derived in Eq.\ (\ref{eq:componentdescription}) for the Markovian form.

From this description it can be seen that the block matrices of \(\mathfrak{Z}_{2 1} \mathfrak{Z}_{2 2}\) are
block-diagonal matrices given by,
\begin{equation}
(\mathfrak{Z}_{2 1}\mathfrak{Z}_{2 2})^{\alpha}_{\alpha} = \int D\tau \Diag{(d\Delta^{\alpha})}S 
\end{equation}
and their components are as given by the description provided in Eq.\ (\ref{eq:componentdescription}).

As \(\mathfrak{Z}_{2 1}\) and \(\mathfrak{Z}_{2 2}\) are block-diagonal traversing their block matrix structure can be 
done with time complexity \(O(N)\) 
with each block containing \(N^2\) components. 

The way this can be accomplished is seen from the block matrix structure of the components presented in Eq.\ (\ref{eq:Zi}) and
(\ref{eq:Xi}).
For us to be able to seek the values of \(D\) and \(\tau\) in a parallel manner we reserve
memory to store the values of these operators for each transition \(\alpha, \lambda\) in Liouville space, for each Bohr
frequency. 
This results in a registry of \(N^4\) many values, denoted by {\tt TauREG}. For the scalar
function \(D\) we have a registry of \(N^2\), denoted by {\tt DREG}. Together, these results are stored in the
registry {\tt DTauREG}
 
This is demonstrated by the following code snippet.
\begin{verbatim}
Z21[] = 0 + 0*i
for block-diagonal matrix beta:
  for column lambda do:
    D = DREG[beta,lambda]
    Dtau = DTauREG[beta,lambda,:,:]
    for row alpha do:
      row = N*(i-1)+alpha
      col = N*(i-1)+lambda
      Zi1[row, col] = Dtau[alpha,lambda]
  end
end
\end{verbatim}
Similarly, \(\mathfrak{Z}_{22}\) is compiled, but for the value of \(S = \pi F \tau^{\dagger}\) we refer to the Fermi
distribution at the Bohr-frequency given by the Dirac measure.
\begin{verbatim} 
Z22[] = 0 + 0*i  
for block-diagonal matrix beta:
      for row lambda do:
         E = H[beta,beta]-H[lambda,lambda]
         Tau = Tau_REG[beta,lambda,:,:]
         S = pi*(1-Fermi(E))*ConjgTransp(Tau)
         for column sigma do:
            Zi2[lambda+(beta-1)*N,
                sigma+(beta-1)*NEEM_Trun]
                = S[lambda,sigma]
         end
      end
   end
\end{verbatim}

In the case of the factors \(\mathfrak{Z}_{3 1}\) and \(\mathfrak{Z}_{3 2}\) there are  \(N^2\) block matrices, each of
them diagonal so they can be compiled with time complexity \(O(N^3)\) as in the case above. Here, we obtain a non-trivial
relation of indices corresponding to a mixed factor in Eq.\ (\ref{eq:componentdescription}).

\begin{verbatim}
Z31[] = 0 + 0*i
for block-line lambda do:
  for block-column alpha do:
    DTau = D_REG[lambda,alpha]
           *Tau_REG[lambda,alpha,:,:]
    for component beta do
      Z31[alpha+(beta-1)*N,alpha+(lambda-1)*N]
                     =DTau[lambda,beta]
    end
  end
end
\end{verbatim}

As before for constructing the factor representing a coupling between the leads and the central system, the
corresponding value, as given by the Markovian form, is sought via the Fermi distribution. 

\begin{verbatim}
Z32[] = 0 + 0*i
for block-line lambda do:
  for block-column alpha do:
     E = H[lambda,lambda]-H[alpha,alpha]
     Tau = Tau_REG[lambda,alpha,:,:]
     R = pi*F(E)*ConjgTransp(Temp_Rtr)
     for block-component sigma do:
        Z32[alpha+(lambda-1)*N,
            alpha+(sigma-1)*N]
                     =R[sigma,lambda]
     end
  end
end
\end{verbatim}
Other dissipative factors present mixed products of the sparse matrix representations described above. The derivation
that we have given here shows a true advantage of our mathematical treatment described in the paper. Furthermore, as
these factors are derived algebraically they can be expressed in terms of their physical variables which is
theoretically appealing for sakes of completeness.

\begin{acknowledgments}
This work was financially supported by the Research Fund of the University of Iceland,
The Icelandic Research Fund, grant no.\ 163082-051, 
and the Icelandic Instruments Fund. We also acknowledge support from the computational 
facilities of the Nordic High Performance Computing (NHPC), and the Nordic network
NANOCONTROL, project no.\ P-13053. HSG acknowledges support from MOST, Taiwan, under grant no.\
103-2112-M-002-003-MY3.
\end{acknowledgments}

\bibliographystyle{apsrev4-1}
%\bibliography{mod_qd}

%%%%%%%%%%%%%%%%%%%%%%%%%%%%%%%%%%%%%%%%%%%%%%%%%%%%%%%%%%%%%%%%%%%%%%
%
\end{document}